# An overview of VANET vehicular networks

Ali Hozouri
Master's student, Department of Computer Engineering, Islamic Azad University, Ardabil branch, Ardabil, Iran

Abbas Mirzaei
Department of Computer Engineering, Islamic Azad University, Ardabil Branch, Ardabil, Iran

Shiva RazaghZadeh
Department of Computer Engineering, Islamic Azad University, Ardabil Branch, Ardabil, Iran

Davoud Yousefi
[1] Department of Computer Engineering, Moghadas Ardabili Institute of Higher Education, Ardabil, Iran

**Abstract**

Today, with the development of intercity and metropolitan roadways and with various cars moving in various directions, there is a greater need than ever for a network to coordinate commutes. Nowadays, people spend a lot of time in their vehicles. Smart automobiles have developed to make that time safer, more effective, more fun, pollution-free, and affordable. However, maintaining the optimum use of resources and addressing rising needs continues to be a challenge given the popularity of vehicle users and the growing diversity of requests for various services. As a result, VANET will require modernized working practices in the future. Modern intelligent transportation management and driver assistance systems are created using cutting-edge communication technology. Vehicular Ad-hoc networks promise to increase transportation effectiveness, accident prevention, and pedestrian comfort by allowing automobiles and road infrastructure to communicate entertainment and traffic information. By constructing thorough frameworks, workflow patterns, and update procedures, including block-chain, artificial intelligence, and SDN (Software Defined Networking), this paper addresses VANET-related technologies, future advances, and related challenges. An overview of the VANET upgrade solution is given in this document in order to handle potential future problems.

Keywords: VANET, vehicle network, artificial intelligence, Block chain, SDN, Ad-Hoc networks

1. Introduction

More than 50% of the world's population currently lives in cities, and this percentage is projected to rise. According to forecasting organizations, this proportion will represent almost 66% of the global population. In terms of road safety, 92% of accidents are frequently attributed to errors in human judgment (such as driver inattention, insufficient supervision, and distraction) and human decision-making errors (such as driving too fast, reacting slowly, safety distance, and misjudging) [1]. In addition, numerous people lose their lives in traffic accidents every year, despite the development of numerous safety-oriented procedures in cars such as anti-lock braking systems (ABS), seat belts, airbags, and rear-view cameras [2]. Therefore, establishing a vehicle communication network that enables vehicles to communicate with the roadside infrastructure and share their data between them to avoid traffic congestion and lead to the realization of an intelligent transportation system is urgently required if there is to be a reduction in vehicle accidents and ongoing optimization of the transportation system. These networks, known as vehicular ad hoc networks (VANETs), are designed to offer low-latency infotainment services and safety-related alerts to both drivers and driverless vehicles [1], [2]. VANET can offer a range of services, including traffic control, entertainment, safety applications, driving assistance, collision avoidance, and safety services. However, due to traffic problems that could have a deadly outcome and cause considerable delays for travelers, VANET developers gave the transmission of vital safety-related information the highest priority [2], [3].

Modern vehicles that are VANET nodes can now offer high computation, storage, sensor, and networking capabilities due to developments in vehicular networks. In a VANET, each vehicle is a potent node that, when grouped together as a cluster, can function as a vast farm of moving supercomputers. Due to this unique characteristic, cars are encouraged to be used not only as mobile machines but also as VANET clouds to utilize their resources and share information. The outcome is a completely new kind of ad hoc vehicular network known as VANET Cloud, in which any vehicles that wish to share their resources must cooperate and join together to share their resources [4], [11].

Initially, on-board units (OBUs) are integrated inside vehicles to enable them to interact directly with roadside fixed units (RSUs) using wireless V2I/V2R (vehicle-to-infrastructure) or V2V (vehicle-to-vehicle) communications. Currently, the vehicle industry is growing from the standpoint of enabling cars to share and exchange data with pedestrians, bicyclists, ground stations (GN), unmanned aerial vehicles (UAV), and other supports through vehicle communication everything (V2X). A communication system that allows vehicles to connect and communicate with their surroundings [3]. In comparison to other Ad-Hoc networks currently in use, such as MANET (Mobile Ad-Hoc Network), FANET (Flight Ad-Hoc Network), and WSN (Wireless Sensor Networks), VANET has a number of distinctive characteristics, including high vehicle mobility, dependence on transportation infrastructure, dynamic network switching, and sporadic network connectivity. With regard to data transfer and communication dependability, these traits cause a number of problems and difficulties [3]. Many scientists and researchers are now looking into how VANET can work with other networks and infrastructures, including cellular networks, satellite networks, WSN, cognitive radio, and most recently UAV networks, as a result of the requirement for reliable communication with low latency and significant data rates in VANET networks [3], [12].

A. Motivation:

Expanding VANET implementation is thought to be essential to the success of many ITS services, which call for considerable advancements in terms of data delivery latency, steady network performance, stability, scalability, and flexibility. Collaboration between VANET and other network technologies, including cellular networks and WSNs, is necessary for this. In order to ensure that electric and autonomous vehicles benefit from ITS services, the rapidly developing automotive industry must take into account several VANET network and vehicle behavior-related difficulties.

Finding the appropriate solutions for these restrictions and gaps can be accomplished by thoroughly examining methodologies and suggestions to identify their flaws and gaps. The issues facing vehicle networks, as well as several technologies connected to the VANET network and communication, are examined in this article.

B. related works:

Studies on VANET are becoming more prevalent today. However, it largely concentrates on particular research fields, like data routing, VANET security, SDN-based VANET, and data dissemination. Few studies on VANET offer a thorough analysis of VANET-related topics, including the effects of VANET problems on various ITS services, wireless technologies that can enhance VANET communications, and the significance of new technologies in advancing VANET networks.

C. Structure of the article

The acronyms used in the article are listed in Table 1. The eight primary sections of this survey cover the following topics.

1. Introduction: In order to give readers and other researchers a new research perspective, this section explains the specifics and motivations behind the production of this study. This section also provides an overview of the article's structure.
2. The concept of VANET: The second section of this article examines the fundamentals of VANET networks, including their components, communication, differentiating characteristics, and smart car components.
3. Ad-hoc network classification: In this section, a quick examination and analysis of the many features and uses of other wireless ad-hoc networks, such as FANET, SANET, TANET, and WSN, is presented with regard to communication, mobility, stability, and other concerns.
4. Wireless access standards are compiled in one place in this section for VANET networks.
5. VANET applications and services: In Section 5, a thorough classification of the VANET applications and services is provided.
6. VANET Challenges: Through a variety of viewpoints and application capabilities, the sixth section informs readers about the difficulties that VANET deployment faces.
7. Connection of VANET with other technologies: In this part, the relationship between VANET and other technologies, including software-defined networks (SDN), blockchain, and artificial intelligence, is investigated.
8. Future research directions: Future research directions are introduced in this section.
9. The study's findings will be provided in the ninth section.

Table 1. A list of the article's abbreviations

| AHS | Automated Highway Systems | AR | Augmented Reality |
|---|---|---|---|
| API | Application Programming Interface | BTS | Base Transceiver Station |
| CC | Cloud Computing | C2C | Container-to-Container |
| CRL | Customer Revocation List | CDN | Content Delivery Network |
| CaaS | Communications as a Service | CA | Certificate Authority |
| DMN | Detection of Malicious Nodes | DOS | Disk Operating System |
| DSRC | Dedicated Short Range Communication | DDOS | Distributed Denial of Service |
| EBS | Electronic Break Lock System | ECC | Edge cloud Computing |
| EC | Edge Computing | FC | Fog Computing |
| FeRAN | Fog-Edge Resource Allocation Network | GPS | Global Positioning System |
| HCI | Human Computer Interaction | HAP | High Altitude Platform |
| HOV | High Occupancy Vehicle | IoT | Internet of Things |
| IoV | Internet of Vehicles | ITS | Intelligent Transportation Systems |
| IaaS | Infrastructure as a Service | IP | Internal Protocol |
| LRPON | Long-reach Passive Optical Network | INaaS | Infotainment as a Service |
| LTE | Infotainment as a Service | LiDAR | Light Detection and Ranging |
| MANET | Mobile Ad Hoc Networks | MAC | Media Access Control |
| MCC | Mobile Cloud Computing | NaaS | Networks as a Service |
| NLOS | Non Line of Sight | OBU | On Board Unit |
| ONF | Open Networking Foundation | PIL | Packet Inter Loss |
| PIR | Packet Inter Reception | P2P | Peer to Peer |
| PCF | Packet Centric Forwarding | PLC | Programmable Logic Controller |
| PDR | Packet Delivery Ratio | PaaS | Platform as a Service |
| PKI | Public Key Infrastructure | QoS | Quality of service |
| QOE | Quality of experience | WSN | Wireless sensor Network |
| RSUC | Roadside Unit Centric | RAN | Rainforest Action Network |
| RSU | Road Side Unit | SUMO | Simulation of Urban Satisfactory |
| SDN | Software Defined Networks | SaaS | Software as a Service |
| STaaS | Storage as a Service | SIOT | Social Internet of Things |
| SIoV | Social Internet of Vehicles | V2V | Vehicle to Vehicle |
| V2I/V2R | Vehicle to Interface/RSU | VRSU | Vehicle to Road Side Unit |
| VCC | Vehicular Cloud Computing | VANET | Vehicular Ad Hoc Network |
| VPKI | Vehicular Public Key Infrastructure | V2RC | Vehicle to Roadside Communication |
| V2X | Vehicles-to-Everything | ABS | anti-locking braking system |
| UAV | unmanned aerial vehicle | FANET | Flying Ad hoc Network |
| RANET | Robot Ad hoc Network | TANET | Train ad hoc Network |
| SANET | ship ad hoc Network | | |

## 2. Concept of VANET
### 2.1. Main components of VANET

A subclass of mobile ad hoc networks (MANET), known as a vehicular ad hoc network (VANET), facilitates communication between nearby vehicles and between vehicles and infrastructure. Roadside units (RSUs) and intelligent vehicles with on-board units (OBUs) make up a VANET's fundamental building blocks. OBUs are hardware elements that are put in every vehicle and allow communication with RSUs and other OBUs. Each car has an OBU installed that receives messages from a source (a vehicle or sensor), verifies them, and then broadcasts them to other vehicles via the DSRC system [5], [13].

#### 2.1.1. The main components of smart cars

Several intelligent technologies have been installed inside of cars to offer real-time measuring and safety services [14]. To form a sensor network, all of these devices are capable of self-organizing communication. Car environmental data is gathered by sensors inside the vehicle. The sensor level collects and stores information such as sensor inputs, navigation services, temperature, behavior identification, and photos for further analysis. The car may independently accomplish autonomous driving control based on this information, including environmental sensing, centralized decision-making, and mechanical control. To aid drivers in preventing accidents, a system architecture ought to be created. The smart car is typically fitted with multi-beam LIDARs, microwave radars, high-resolution cameras, etc. to acquire accurate and sufficient environmental information [15]. An intelligent vehicle typically includes the following components and technology, as depicted in Figure 1 [6]:

1. CPU: Effectively computes instructions by carrying out arithmetic, logical, and I/O operations.
2. Wireless transceiver: uses vehicle-to-vehicle and vehicle-to-infrastructure connections to transmit data and information.
3. GPS receiver: This device receives data from the Global Positioning System and supports the provision of navigational services. A GPS may automatically report the precise location of the vehicle even with an accuracy of less than 1 meter by integrating with communication devices. The information gathered makes it easier to create groups and can quickly and accurately warn others about accidents, unsafe driving conditions, etc.
4. Sensors: A car has a number of sensors inside and outside to measure things like speed, separation from other vehicles, and other things. An ultrasonic sensor, for instance, that relies on sound waves being reflected off of objects. It determines the proximity and velocities of surrounding objects.
5. I/O interface: enables simple human-vehicle communication.

6. Radar: Uses radio distance to locate and keep track of nearby vehicles. Long-range radars and short-range radars are the two types of vehicle radars. At the moment, adaptive cruise control systems utilize these sensors.
7. LIDAR: light detection and ranging with the help of a sensor with one-dimensional scanning capabilities, a vehicle's relative distance can be precisely determined by scanning a horizontal surface with laser beams. This sensor transmits infrared light pulses with a wavelength of 850 to 950 nm using powerful laser beams.
8. OBU: Internal unit. Each vehicle in the system is fitted with an OBU, which regulates vehicle connection with RSU, SMBS, and other cars through DSRC/LTE-V.
9. LCS: Local camera sensor is a sensor that tracks the actions of the driver while simultaneously reliably and accurately detecting objects.

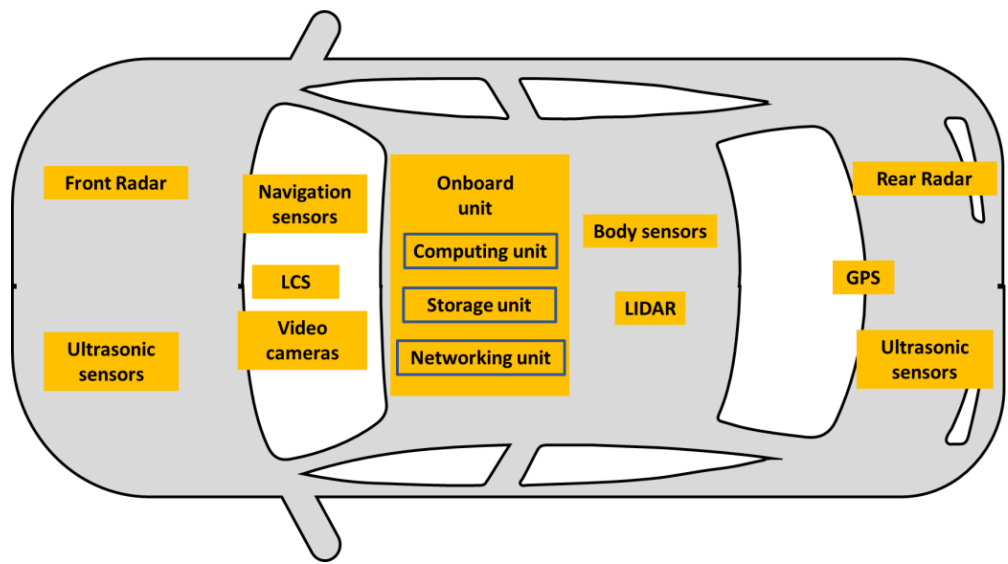

Figure 1. Components of a smart vehicle

## 2.2. VANET communications

The development of intelligent transportation systems, or VANETs, has become increasingly necessary as a result of improvements in public transportation networks worldwide, whether for business or personal use [3]. Real-time information about the state of the roads or data pertaining to safety can be wirelessly transferred between vehicles in VANETs or by other channels, including RSUs, UAVs, etc. Roadblocks and traffic accidents are easily preventable [3]. Additionally, VANETs can share additional entertainment content like news, games, and Internet access, which can make lengthy commutes enjoyable. As a result, VANET enhances the driving experience while lowering accidents and congestion [3], [16].

Vehicles must be upgraded with sensors, navigation systems like GPS, multimedia devices, and wireless modules in order to join VANET. To avoid accidents and unexpected mishaps, sensors and multimedia techniques can be utilized to feel the surroundings and recognize items around vehicles, such as other vehicles, barriers, and passengers. In contrast, wireless communication modules offer a variety of communication linkages that can be classified in accordance with the communication entities shown in Figure 2 [7].

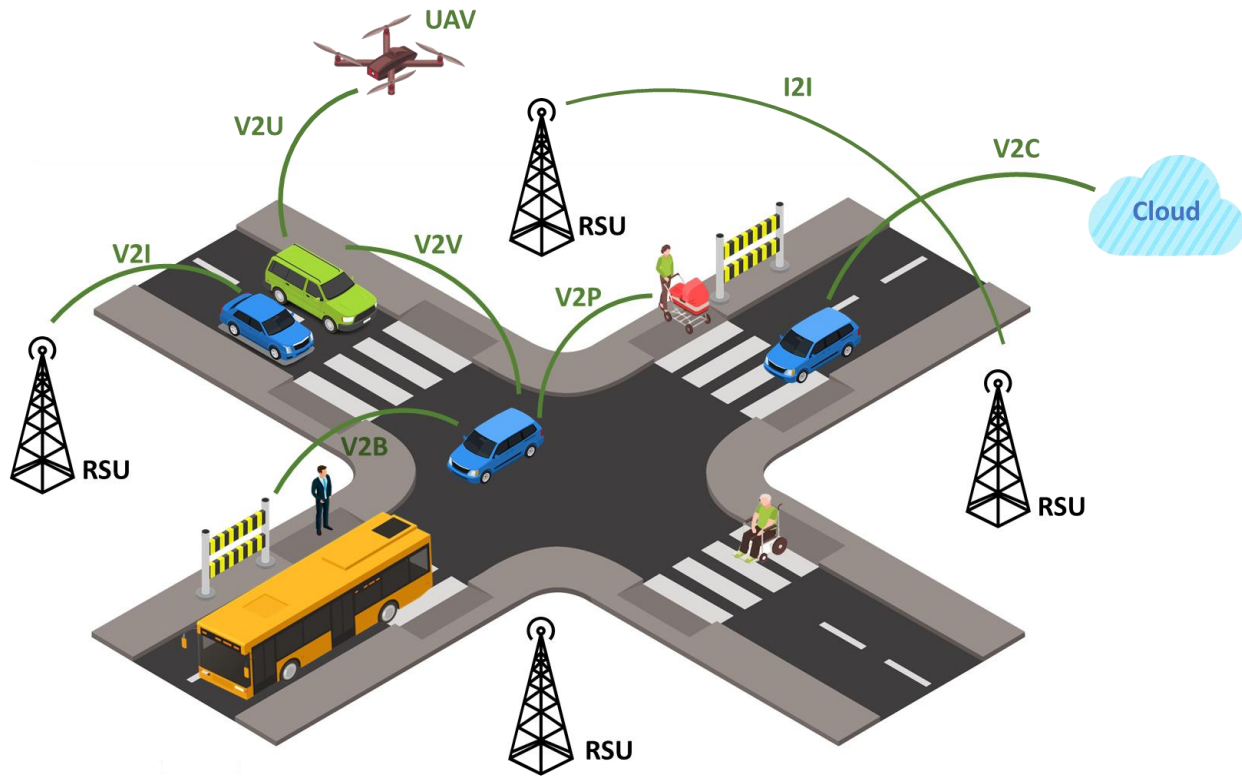

Figure 2. Basic communication of VANET.

The various forms of VANET communication [3] include:

1) Without the need of infrastructure, vehicle-to-vehicle (V2V) communications take place directly between the vehicles. Sharing safety-related information is the major purpose of these connections.
2) Communication between automobiles and roadside infrastructure, such as roadside units (RSU) or cellular base stations, is made possible by vehicle-to-infrastructure (V2I/V2R) communications. These connections can be used to share announcements and give access to the Internet.
3) Road traffic data is shared between infrastructures in certain locations via infrastructure-to-infrastructure (I2I) connectivity.
4) Data sharing between portable devices and moving cars is done through vehicle-to-pedestrian (V2P) communication.
5) Data sharing with roadside barriers requires vehicle-to-barrier (V2B) connectivity. To reduce traffic accidents, this communication makes use of the numerous sensors in automobiles.
6) RSUs and cloud servers can connect via vehicle-to-cloud (V2C) technology for a variety of purposes, including data processing, decision-making, and transportation forecasting.
7) Vehicle-to-Urone (V2U) communication uses ground-to-air links to wirelessly transport data between cars and drones.
8) Vehicle-to-Sensor (V2S) communication allows embedded sensors in cars to communicate information about the road and its surroundings.

A very large wireless network is known as the VANET. The primary causes of the dynamic VANET network topology are high vehicle mobility and a diversified dispersion of routes. In order to achieve the best outcomes, VANET communications must adjust to the dynamic nature of VANETs and the various QoS specifications for the services offered. On the other side, many applications have numerous QoS needs. For instance, whereas non-safety applications require high-throughput networks, safety-related services must be accessible with low latency and good dependability [3], [17].

2.3. VANET's distinctive characteristics

VANETs have traits and difficulties that make service delivery and confidence in their dependability complex and multi-metric issues. The following is a description of a few of these characteristics:

1. Mobility variation: In a VANET, communication entities might be stationary (like RSUs), slowly moving (like cars in traffic and at intersections), or quickly moving (like cars on dispersed highways and roads). The variance in nodes' comprehension makes communication within the VANET difficult. [3], [1].
2. Limitation of motion: The infrastructure of the public transportation system limits the movement of vehicles in a VANET environment. City to city differs in the distribution of its highways. Road types also differ depending on the region they are located in [3], [18].
3. Frequent network fragmentation: Location and timing may affect the number of vehicles. VANET network fragmentation will be significantly impacted by the density and mobility of VANET nodes. By decreasing node density, the network will become fragmented, which will have an adverse effect on the flow of communication and packet delivery. Furthermore, the topology of the VANET is more dynamic when cars are moving at high speeds. It is possible to partition the VANET network into various parts as a result [3], [19]
4. Heterogeneity: A VANET comprises of numerous nodes with various properties and functions. These nodes could be stationary like RSUs or mobile like vehicles. Additionally, certain vehicles need the communication of entertainment information, while others require the exchange of safety-related information. A VANET system deals with the heterogeneity of nodes and services [3], [20].
5. Scalability: In the VANET, connections may span small cities, numerous towns, or big cities and countries. As a result, the VANET network's geographic scope is virtually limitless. In automotive communications, scalability is one of the major issues. This implies that even with a small number of nodes to transmit, VANET connection performance should be dependable. Today, using UAVs for direct data relay or storage is a fresh, workable approach to VANET scaling problems [3], [21].
6. Unlimited power and computing resources: The communication between nodes in the VANET network is not restricted by power or storage. VANET eliminates power and processing resource issues by embedding OBUs in vehicles that operate on continuous and unlimited energy sources from vehicle batteries. The use of different energy-saving strategies, including RSA and ECDSA approaches, is supported [3].
7. Spectrum Scarcity [3]: Wireless technology standards for automobile networks include wireless access in vehicle environments (WAVE) and dedicated short-range communications (DSRC). In large-scale dense vehicular networks, reliability and scalability problems with DSRC-based VANETs have been observed in recent research studies. The dependability and scalability of insufficient and scarce frequency channels for automotive communication systems is one of the primary causes of these issues. Each vehicle periodically uses seven DSRC channels for data exchange. This implies that all vehicles must make an effort to transmit vehicle data packets on these channels. As a result, this channel competition may cause network congestion, increased packet delay, and decreased throughput, which would lower the quality of service as a whole.

8. Environmental Effects: In a VANET, all communication occurs outside, where the environment's impact on electromagnetic signals will be comparatively high. Buildings, cars, trees, and other objects can interfere with these signals as they travel through the air, resulting in a variety of signal disturbances include multipath propagation, channel fading, and signal shadowing. Additionally, climate change has an impact on VANETs' high-speed, low-latency data connections [3].
9. Accuracy of information [3]: Positional information from navigation systems like GPS and GNSS is used by several data delivery mechanisms in the VANET environment. Unfortunately, these devices are unable to offer precise location data. Additionally, the troposphere's and ionospheres atmospheric effects may weaken it. In a densely populated metropolitan location, the GPS signal's accuracy is typically between 5 and 30 meters under ideal conditions. This precision is negligible and could have a negative impact on the dependability of low-latency VANET services. In contrast, RSU gathers information about the present cars in its coverage area. The cloud or SDN may receive this data in order to analyze it and make subsequent judgments. This information may not be accurate and may need to be changed due to the rapid pace and constant movement, which could lead to mistakes in analysis and good judgment.
10. Fault Tolerance: The main requirement for a variety of safety-related services in VANETs is real-time communication. Any inaccuracies, however, could result in further delays in the distribution of the data, which would generate terrible traffic jams and calamities. Preventive security measures must be put in place as soon as possible to avoid such occurrences and to give drivers the ability to act quickly in an emergency [3].
11. Data Security [3]: Data must be presented in a secure manner for communication-management to be efficient and trustworthy. The location of the data must guarantee that the packets are not altered while being transmitted. Additionally, it must be encrypted so that no unauthorized party may decipher its contents. Last but not least, confirm that the sender and not a third party sent it. Security is a crucial concern in the adoption of the VANET since, in the event of a hack, vehicles could surely be controlled by hackers, causing traffic jams and potentially deadly calamities. Data security and non-repudiation, however, are crucial difficulties in VANET adoption because of VANET characteristics including dynamic network topology and vehicle mobility. Additionally, keys are essential components of the cryptographic methods used in a VANET context to encrypt and decrypt sensitive data. Key management is extremely difficult due to the VANET network topology's frequent alterations. Any strategy to strengthen VANET security, however, must take into account how keys are created and delivered. Revocation, or the process of disregarding the keys of harmful subscribers, is one of the main elements of key management. As a result, the VANET network topology extension may provide a lengthy list of reversible keys, adding to the complexity of the key revocation process.
12. Data privacy: Many consumers do not want their car information to be shared or utilized to help them locate their destination. Data privacy is becoming one of the most crucial issues that must be balanced between the public's desire and high levels of privacy [3].

3. **Classification of MANET networks**

The several types of MANET wireless networks include wireless sensor networks (WSN), robot ad-hoc networks (RANET), flight ad-hoc networks (FANET), train ad-hoc networks (TANET), and ship ad-hoc networks (SANET). Figure 3 illustrates how this classification is based on the categories of communication entities and the environments in which they are deployed [8]. But each of these operates with a number of worries and has unique traits, problems, and obstacles. WSNs are a collection of small, cheap sensing gadgets that can gather information from a close-by location and transfer it immediately to a central processing unit called a sink. There are many industrial, domestic, and civic applications for WSNs. However, the static or sedentary properties that set WSN nodes apart from each other mean that their density is typically low. One of the major difficulties in deploying WSNs is the power supply at such nodes [3]. FANET networks are made up of a collection of UAVs, and at least one of them needs to be able to connect with satellites or ground-based stations (GBS). Through wireless connections between the drones, the drones assembled in a given FANET can work with one another. In order to coordinate, manage, and communicate with UAVs, FANETs typically need more sophisticated hardware and software systems. To prevent collisions in FANET networks, effective node coordination methods are required [3].

Robots located in a certain area make up RANET networks. To optimize their performance or prevent collisions with objects and other robots, these robots must interchange sensor data with other systems or centralized units. Robots can move intelligently, although they typically have limited mobility. Additionally, robots use cognitive management to control their limited energy capacity [3].

Wide-area networks called SANETs are utilized to cover marine communications between ships [3]. These networks experience signal propagation delays, frequent network disconnections, and low node densities.

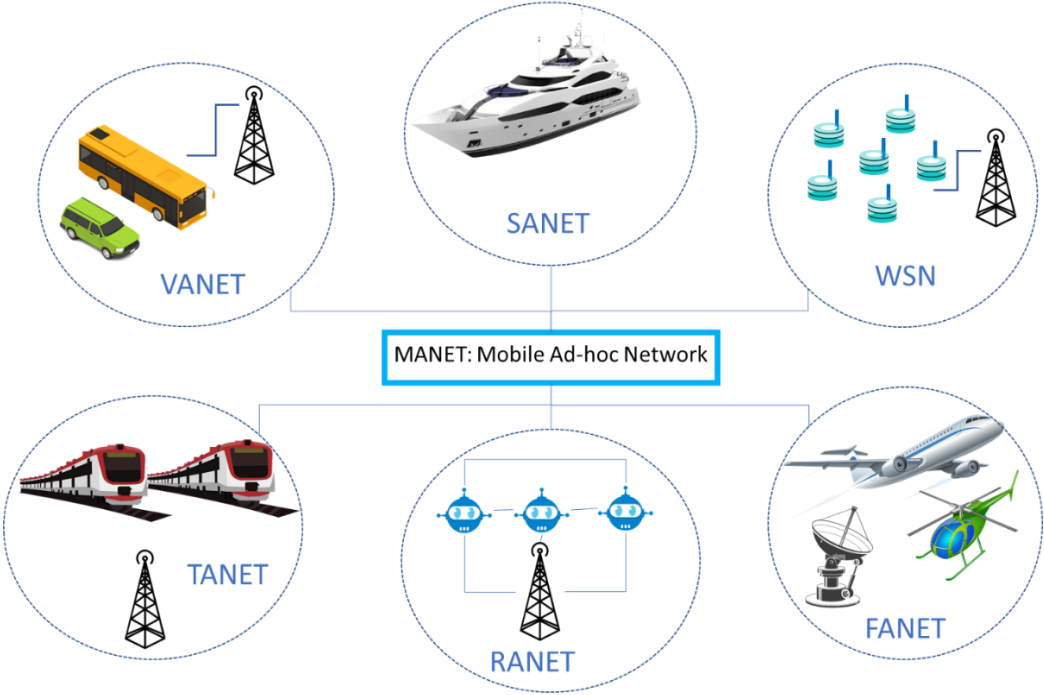

Figure 3. Different classes of MANET network.

## 4. VANET wireless access standards

VANET exhibits significant heterogeneity due to the collaboration of several communication and information technologies. Inter-vehicle, intra-vehicle, and infrastructure-vehicle communication are only a few of the different types of communication that exist. In the beginning, VANETs mostly relied on DSRC to offer vehicular communication. Researchers came up with more thorough solutions as a result of the standard's limited capacity and the significant likelihood of traffic jams and extended delays. Additionally, the expansion of VANET services and the requirement to maintain the continuity and scalability of VANET connections drive the adoption of various wireless communications technologies, including cellular communications (4G/LTE), communications over Wi-Fi, and communications over Bluetooth. It makes possible short-range static (Zigbee, Bluetooth, and Wi-Fi), ultra-reliable and low-latency communications (5G, MmWave, VLC), satellite communications, and cognitive radio communications [3].

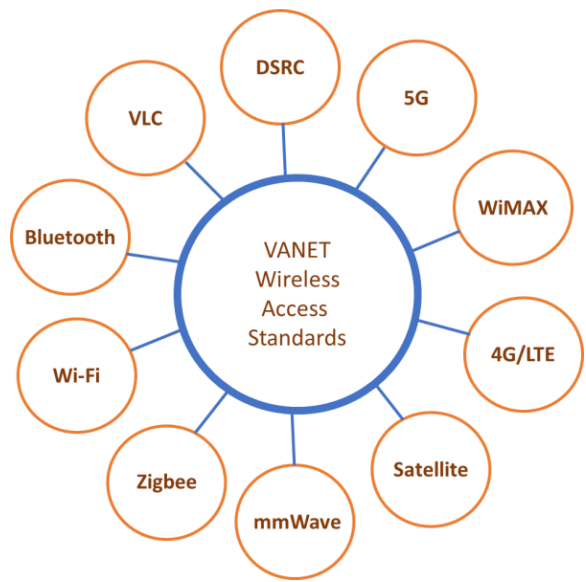

Figure 4. Wireless access standards used in VANET communications.

## 5. VANET applications and services

Universities and businesses have created several VANET services and applications as a result of the increased demand for ITS adoption in smart cities. Therefore, VANET may offer a wide range of services and applications to governmental organizations as well as motorists and passengers. The use of these services and applications can be categorized into applications for safety, entertainment, traffic improvement, and monitoring of the driving system [3], [22]. Numerous applications fall within each of these areas. The categorization of VANET services and the many applications within each category are shown in Figure 5.

### 5.1. Safety related applications

The main functions of VANETs are safety-related applications, which are offered to prevent and reduce the frequency of traffic accidents. To prevent or lessen the effects of traffic accidents, safety-related data should be collected, transferred, and proactively provided with as little delay as possible [3]. Driver assistance applications, safety information delivery applications, and driver alert applications are three categories into which these applications can be divided. Information concerning intersections, such as traffic light status and timing, traffic volume, road priority, and the present weather conditions, may be included in warning messages that are sent to moving cars in advance to help them take the proper measures [3]. Because of this, fewer people are killed in auto accidents.

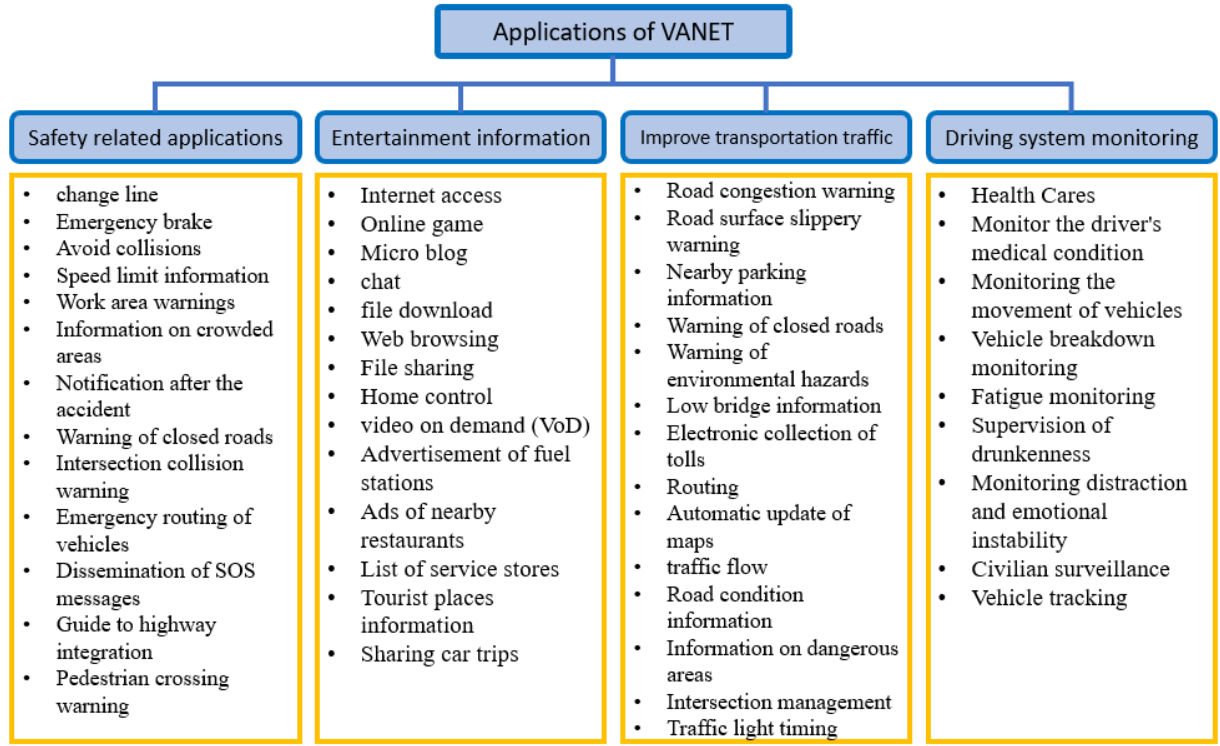

Figure 5. Classification of VANET services

5.2. Infotainment applications

Drivers and passengers can be entertained while traveling with infotainment applications, which include VoIP, online gaming, video sharing, and other non-safety-related activities. These applications can handle delays and may demand high bandwidth. However, this form of application can typically be divided into three categories: urban announcement applications, online e-commerce applications, and entertainment applications [3]. Passengers can utilize the VANET Internet connection if other well-known Internet access networks (Wi-Fi, WIMAX, etc.) are not available. Even with such networks, a car that is connected to the Internet through one of them can give other cars access to the Internet via a VANET. Several businesses use VANET to promote their products and services by disseminating commercial messages via the local VANET infrastructure [3], [23].

5.3. applications for improving traffic in transportation

By regulating traffic flow and reducing congestion, traffic improvement initiatives aim to enhance road traffic and lower the number of accidents on the road. Vehicles are given traffic situation information; if there is a chance to escape traffic to shorten travel time, drivers may select an alternative route [3].

Intersection management, traffic congestion management, road condition management and transportation information programs are the three primary areas under which transportation traffic improvement initiatives fall.

### 5.4. Driving system monitoring programs

These systems deal with monitoring the driver's health while driving as well as the state of the vehicle and its numerous components. Numerous categories can be used to categorize these applications, including "health care applications," "driver physiological behavior monitoring applications," "vehicle motion monitoring applications," and "vehicle mechanics monitoring applications" [3], [24].

## 6. VANET's challenges

There are a number of problems that VANET may experience, which can severely restrict its utility. Based on VANET applications, data networks, and VANET resource management, Figure 6 illustrates how VANET issues are categorized. Researchers and developers can use this classification to identify areas that need and offer opportunities for improvement [3].

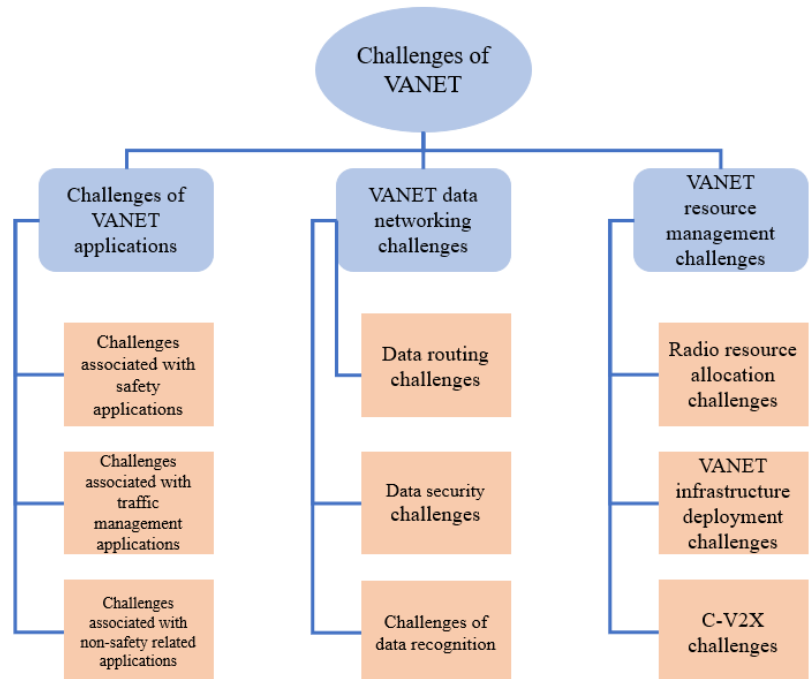

Figure 6. [3] Challenges of VANET.

### 6.1. The challenge of applications

Offering proactive warning messages with prompt responses and excellent dependability to ensure safe travel is one of VANET's most favored objectives. However, there are numerous obstacles to applying and deploying VANET services and applications, and these obstacles vary depending on the service's needs [3], [25].

### 6.2. Data networking challenges

Inter-vehicular communication (IVC) is the foundation of VANET communication, which enables vehicles to build mobile communication networks and share their data. This eliminates the need for centralized communication systems, allowing ITS to offer a variety of services. The three key components of data routing, data security, and message propagation are the fundamental issues that data networks in vehicular communication must overcome [26].

### 6.3. VANET resource management challenges

Many resources, such as bandwidth channels and RSUs, are shared by shared vehicles in a VANET infrastructure. The deployment of VANET is faced with several challenges as a result of this sharing [3].

### 6.4. Radio resource allocation challenges

Techniques for medium access control (MAC) are essential for guaranteeing that all vehicles have fair channel access while reducing data packet loss. In order to increase MAC performance in VANET, additional work will be required due to the primary characteristics of VANET, such as high dynamics, intermittent connectivity, diverse QoS needs, and security difficulties. There are still a number of MAC research issues and open questions that need to be resolved before VANETs can handle secure and non-secure services, despite efforts to make the MAC protocol perform better in VANETs [27].

## 7. Connection of VANET with other emerging technologies
### 7.1. Software Defined Networks (SDN) for VANET

SDN is the separation of system (control plane) and transmission capacities (data plane). Before transferring data to a network device, tasks are formed in the controller, and controllers specify the logic that governs network behavior [9]. SDN is defined as a network that can be utilized to overcome the restrictions of old traditional networks that have been used for decades [9]. The following are the VANET abstractions that can be used to integrate with SDN concepts:

- 7.1.1. Control Screen: The control screen is used to control all of the cars and RSUs on the data screen. It saves and executes the status of all SDN switches, as well as collecting data for each car, such as vehicle position, speed, and network connectivity. The Global Positioning System (GPS) is a requirement for the network topological data that was gathered. Such information can be used by a controller to make routing decisions and then choose the best route for sending data packets to their destination [9].
- 7.1.2. Data Plane: To create this plane, connectivity-providing network elements were used. Vehicles and RSUs with SDN/OpenFlow (OF) switches are examples of network components. While RSUs are fixed components of the data sheet, vehicles are regarded as its fundamental components. Different deployment procedures are applied by SDN switches. The control interface is used to configure each vehicle. To optimize the network setup, the SDN controller receives notifications from each vehicle. For V2V and V2R/R2V communication in vehicle networks, IEEE 802.11p is employed. The OF flow table in the newly built RSU is where all vehicle information, including direction, location, and speed, are placed [9].
- 7.1.3. OF protocol: OF is typically utilized in the communication interface between the data and control planes of the SDN infrastructure [9] and may be efficiently transmitted and arranged in the terrestrial network. This protocol's three primary components are the OF switch, protocol, and controller, as depicted in Figure 7. The flow table, secure channel, and OF protocol are the three components of an OF switch. A flow entry is present in each flow table and is used to control the flow in the switch. Access between the switch and the controller via the OF protocol is made possible with the help of the security channel. The management of the packets moving through the network is greatly aided by the OF controller. The contention field, timeout, and activity are all included in the flow input. The distance to the RSU and the vehicle's speed can be used to calculate the rest period [9].

7.1.4. SDN in VANET environment: This section explains how SDN networking can improve VANETs. Intelligence and network status can be at the forefront of a VANET due to the separation of the control and data planes. The VANET environment has the potential to be employed in achieving high compatibility, adaptability, and scalability. The control plane and the data plane are the two primary components of SDN. In order to communicate between these two layers, a router protocol is used. SDN governs the network's flexibility and programmability, making the system aware of and responsive to changing conditions and requirements. Awareness enables the software-defined VANET to make optimal judgments depending on the data gathered from the sources [9]. Through an application-defined interface known as the Southbound Interface (SBI), the SDN controller monitors the status of data plane elements and injects forwarding rules. Of is the most well-known SBI. A common programming abstraction is used at the top layers in addition to the controller, i.e., the north interface supports network applications. Expansion of SDN principles (flexibility, scheduling, and centralized control) in network management and communication resources in automotive networks, including channel allocation and network selection optimization, interference reduction in multi-channel and multi-radio environments, and environment-specific packet routing decisions In high-speed scenarios, multi-hops and efficient mobility management are advantageous.

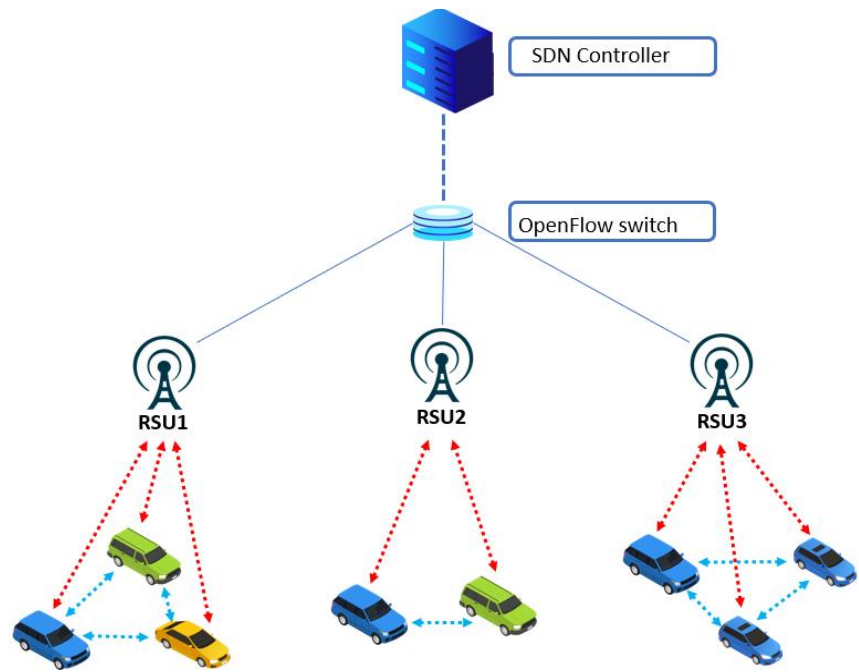

Figure 7. OpenFlow switch in the Internet of Vehicles.

## 7.2. The emergence of AI within the Internet of Vehicles

In an effort to handle complicated and difficult tasks that often require human participation, artificial intelligence is applied in a wide range of industries and applications. In order to solve difficult automotive problems, artificial intelligence is used, which advances the sector's economy and industry. The Internet of Vehicle applications, including speed change, object identification, building trajectory, lane change, real-time navigation, traffic signal control, and video surveillance, have made significant improvements in artificial intelligence that have made autonomous driving possible. Argo AI, Cruise, Waymo, and Aurora are a few examples of businesses spearheading the development of self-driving car systems at the industry level as the field of self-driving research expands quickly. These businesses were able to put theory into practice by using computations on sensed data and radars to pinpoint items, forecast other people's behavior, and plan future pathways up to three football fields away [10].

When a vehicle is in self-driving mode, it gathers the necessary data before it begins to travel on the road. Sending a radar beam continuously for prediction, detection, and decision-making is the simplest way to accomplish this. The car can recognize its position on the road and be able to comprehend the surrounding environment by putting this data into an artificial intelligence model that can use multi-factor reinforcement (RL), game decision theory, or logical reasoning. This is accomplished by artificial intelligence, which recognizes various objects, including the color of traffic lights, moving pedestrians, objects crossing the road, and stationary objects. Artificial intelligence is also employed for instant action planning and execution. Turning left or right, accelerating or decelerating, changing lanes, choosing when to stop when necessary, or halting straight away to prevent an accident are a few examples of decisions. Furthermore, a key model for autonomous driving relies on anticipating the behavior of drivers and nearby objects. In a nutshell, AI helps the Internet of Vehicles by deciding on the best course of action through object detection, decision-making, and behavior analysis [10], [15].

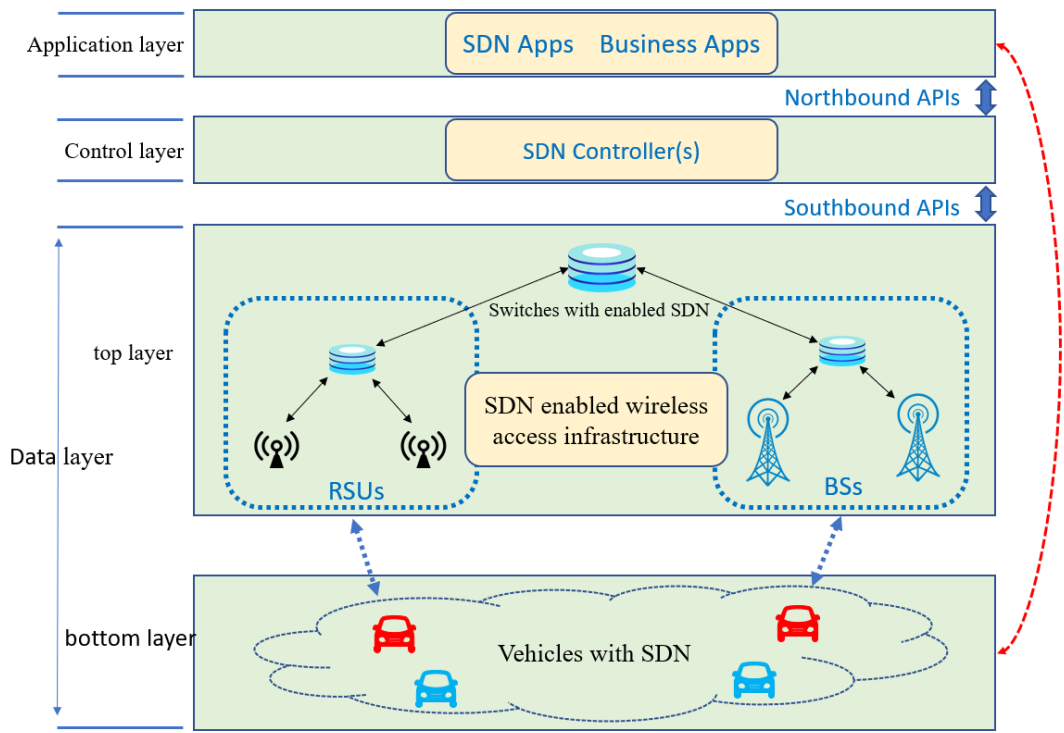

Figure 8. Components of software-defined vehicular networks.

### 7.3. Blockchain to secure the Internet of Vehicles

Parallel to artificial intelligence, blockchain is a decentralized, distributed digital ledger that was created as the underlying network architecture for the well-known, safe cryptosystem "Bitcoin". It has lately been embraced by a variety of applications, including financial services, IoT, voting systems, and healthcare [10]. Without any centralized control, Blockchain runs on a totally distributed peer-to-peer architecture. As a result, applications and architectures that rely on blockchain to manage their network infrastructure benefit from high levels of data accessibility, security, privacy, and trust. Blockchain, for instance, offers transparency by making a complete copy of the database available to all participating nodes. Prior to updating their databases, nodes also need to come to an agreement on new transactions. Blockchain 2.0, which added several features, including the foundation of smart contracts, was published as a result of its broad success. A smart contract is a piece of self-executing code that begins to run whenever its predefined conditions have been satisfied, without the need for any outside authority [10]. Derivative blockchains come in several forms, each of which has a specific function. Examples include:

Public Blockchain[10]: An entirely decentralized ledger that is managed without the need for a centralized entity. These blockchains are open to everyone. Examples of public blockchains that allow anonymous involvement to improve their decentralization are Bitcoin and Ethereum.

Private Blockchain [10]: A permission-based blockchain primarily used by private organizations. It depends on a centralized authority to control who may access the network, preventing some users from reading or writing within it.

Blockchain Consortium [10]: This kind of blockchain is a hybrid one. For its management, it is dependent on a collection of authorized entities. Such a type might use a hybrid access strategy that enables permitted communication with the outside world. Consortia for blockchain technology include Corda and Quorum.

At the industry level, many top automakers have started investigating blockchain to enhance the driving experience by safely transferring data via V2X [10]. For instance, General Motors and BMW are investigating a blockchain-based system to share data about self-driving cars with one another, which will aid in the operation of self-driving cars. Volkswagen, on the other hand, is developing a tracking system that uses blockchain technology to stop dealers from inflating the cost of their vehicles by tricking odometers. For the security and privacy required for the Internet of Vehicles to function, experts have recently concentrated on incorporating blockchain into IoV. To promote road safety and efficiency, certain activities specifically use blockchain to provide data authenticity and trustworthiness. Others concentrated on protecting drivers' privacy from data breaches and online harassment [10].

Several blockchain-related strategies:

Increasing security: A lot of experts have recently become interested in the veracity of communications sent between cars. By determining the veracity of broadcast messages, several of these researchers have enhanced the security of the IoV paradigm using the blockchain approach. To prevent malicious attacks on IoV and to create a secure environment, several researchers have suggested authentication procedures [10].

Privacy: For reasons of privacy, the majority of blockchain solutions demand public key encryption. Blockchain was consequently added to the IoV to improve the privacy of drivers. Finding ways to shield automobiles from the leakage of sensitive personal data was a focus for several researchers. Naturally, the majority of communications utilizing blockchain technology will be encrypted, making it very challenging for outside parties to decode and extract crucial information [10].

Due to the unpredictable and unexpected behavior of automobiles, vehicular Internet applications are fairly complex and call for sophisticated implementation strategies. As a result, scientists now use artificial intelligence to support decision-making. Despite these efforts and contributions, the IoT infrastructure's computing load is further increased by AI, which necessitates a hidden, malicious-free environment to function as intended. Otherwise, AI could become unreliable and endanger road safety because of its susceptibility to cyberattacks. Such attacks have the ability to control these algorithms and disrupt their behaviors with relatively little alteration, leading to erroneous conclusions that can endanger people's lives.

This is primarily caused by the algorithms' unpredictable decisions. In this context, the combination of blockchain and artificial intelligence helps infrastructure management and application development, enabling a decentralized, intelligent, and secure Internet of Vehicles. The vehicular infrastructure may function worse if attempts to use blockchain to bypass the aforementioned restrictions result in a larger computational load. Therefore, in order to successfully use and maintain the Internet of Vehicles infrastructure and satisfy the objectives of both technologies, we need an architecture that blends artificial intelligence and blockchain and makes use of automotive edge computing [10].

8. Issues for future VANETs

Figure 9 provides an introduction to the numerous technologies that might be researched in this area. They have changed over time as a result of extensive study, testing, and deployment, but further testing and development are needed to make them suitable for the VANET services' new challenges and needs [3].

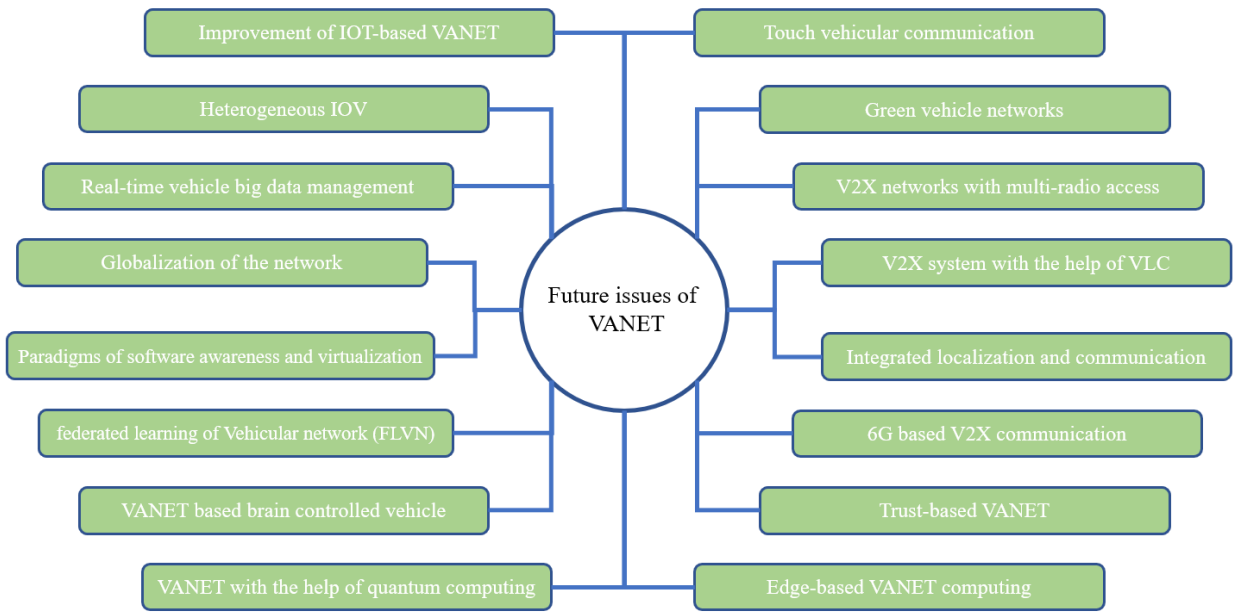

Figure 9. Issues for the future of VANET

9. Result

VANET network improves road transport management and human safety against traffic accidents. Numerous studies have been done to look into possible solutions for the different problems and difficulties that arise in vehicle networks, including routing, security, and communication improvement. However, there aren't many studies that thoroughly examine vehicle networks from several angles, including communication, applications, and problems. As a result, this article offers a thorough analysis of current wireless access technologies that are applicable to VANETs.

To clearly identify their usability and use for applications connected to safety, infotainment, transportation traffic improvement, and driving system monitoring, VANET services are provided with a thorough definition and additional explanation of their classification.

Additionally, the challenges and issues of a vehicle infrastructure are studied and organized from the perspectives of applications (difficulties of safety-related applications, difficulties of traffic management applications, difficulties of non-safety-related applications), data networks (difficulties of data routing, difficulties of data security, difficulties of data dissemination), and VANET resource management (difficulties of radio resource allocation, difficulties of VANET infrastructure deployment, difficulties of C-V2X).

Additionally, future issues with VANET and ITS (intelligent transportation systems) as well as research areas that support the development of novel approaches and solutions by fusing together established and developing technologies are discussed. Future V2X communications will require VANET to be integrated with other new enabling paradigms; as a result, the difficulties of such integrations are examined in this paper. This route also represents a fresh approach for academics to enhance VANET communication patterns, moral extensions, high-performance data distribution, and many other things. Additionally, fresh ideas and methods for fusing such technologies with automotive networks are provided here.

However, while VANET appears promising in contemporary transportation systems, there are a number of issues that need to be resolved, according to the study's conclusions. Although it is utilized in a number of industries, including safety, traffic safety, convenience, entertainment, and health care, further applications and integrations should be made. Overall, looking at a vehicle network from many angles demonstrates that one-size-fits-all concepts may be realized, enabling readers and researchers to locate everything in a single box while encouraging additional researchers to pursue interesting areas of research.